\documentstyle[12pt,twoside,psfig]{article}
\begin{document}

\title{\Large\bf Spintessence: a possible candidate as a driver of 
the late time cosmic acceleration}

\author{Narayan Banerjee\footnote{email: narayan@juphys.ernet.in}~~~and~~
Sudipta Das \\
Relativity and Cosmology Research Centre,\\
Department of Physics, Jadavpur University,\\ Kolkata 700 032,\\India.
}

\maketitle
\vspace{0.5cm}
{\em PACS Nos.: 98.80 Hw}
\vspace{0.5cm}

\pagestyle{myheadings}
\newcommand{\be}{\begin{equation}}
\newcommand{\ee}{\end{equation}}
\newcommand{\bea}{\begin{eqnarray}}
\newcommand{\eea}{\end{eqnarray}}

\begin{abstract}
In this paper, it is shown completely analytically that a spintessence 
model in the dust dominated universe can very well serve the purpose 
of providing an early deceleration and a present day acceleration. 
\end{abstract}

\par The stunning results of the observations on the luminosity - redshift relation of some distant supernovae \cite{perl,gar}, that the universe is currently undergoing an accelerated phase of expansion, poses a serious challenge for the standard big bang cosmology. As the standard gravitating matter gives rise to an attractive field only, this challenge is negotiated in the standard model by invoking some field which gives rise to an effective negative pressure. For many a reason, a dynamical `dark energy' is favoured against the apparently obvious choice of a cosmological constant $\Lambda$ for providing this negative pressure \cite{varun}. This dynamical dark energy is called a quintessence matter. It cannot be overlooked that a successful explanation of the formation of structures in the early matter dominated era crucially requires an effectively attractive gravitational field and thus a decelerated phase of expansion must have been witnessed by an early epoch of matter dominated universe itself. Very recently Padmanabhan and Roy Chowdhury \cite{padma} showed that the data set, only showing the accelerated phase of expansion, can well be interpreted in terms of a decelerated expansion in disguise. The acceleration only becomes meaningful if the full data set shows deceleration upto a certain age of the universe and an acceleration after that (see also ref \cite{amen}). In keeping with such theoretical requirements, actual observations indeed indicate such a shift in the mode of expansion - deceleration upto a higher redshift regime (about $z \sim 1.5$) and acceleration in more recent era, i.e, for lesser values of $z$ \cite{riess}.\\
\par This observation indeed came as a relief, as the formation of galaxies could proceed unhindered in the decelerated expansion phase. It also requires all quintessence models to pass through certain fitness tests, such as the model should exhibit a signature flip of the deceleration parameter $q$ from a positive to a negative value in the matter dominated era itself. Quite a few quintessence models do exhibit such a signature flip of the deceleration parameter. One very attractive model was that of ~`spintessence' proposed by Boyle, Caldwell and Kamionkowski \cite{boyle}. It essentially works with a complex scalar field and a potential, which is a function of the norm of the scalar field. The scalar field, 
\be
\psi = \phi_{1} + i \phi_{2},
\ee

can be written as

\be
\psi = \phi e^{i\omega t},
\ee

i.e, the complex part is taken care of by a phase term. This kind of a scalar 
field is already known in the literature in describing a `cosmic string' 
\cite{vilen} although it has to be noted that a cosmic string has a very 
specific form of the potential $V(\phi)$, whereas, the relevant form has to 
be found out for a quintessence model. Boyle et al discussed the so called 
spintessence model in two limits separately, namely for a high redshift region 
and also for a very low redshift region. Apparently this model gives exactly 
 what is required, an acceleration for the low $z$ limit whereas a deceleration 
for a high redshift limit. In the present investigation, it is shown 
completely analytically that the model indeed works. For high value of the 
scale factor $a$, the deceleration parameter $q$ is negative whereas for a low 
value of $a$, $q$ is positive. The most important feature is that the value 
of $a$, where the sign flip of $q$ takes place, can be analytically expressed 
in terms of the parameters of the theory and constants of integration.

 \par If we take the scalar field as given by equation (2), Einstein's field 
equations become
\bea
3\frac{ \dot{a}^2}{a^2} = \rho + \frac{1}{2} \dot{\phi}^2 + \frac{1}{2} {\omega}^2 {\phi}^2 + V(\phi)\\
2\frac{\ddot{a}}{a} + \frac{ \dot{a}^2}{a^2} = - \frac{1}{2} \dot{\phi}^2 -\frac{1}{2} {\omega}^2 {\phi}^2 + V(\phi),
\eea

where $a$ is the scale factor, $\rho$ is the energy density of matter, $V$ is 
the scalar potential which is a function of amplitude $\phi$ of the scalar field, and the phase $\omega$ is taken to be a constant. An overhead dot implies 
differentiation w.r.t. the cosmic time $t$. In a more general case, 
$\omega$ might have been a function of time. 
\par The matter distribution is taken in the form of dust where the 
thermodynamic pressure $p$ is equal to zero. This is consistent with the 
`matter dominated' epoch. This leads to the first integral of the matter 
conservation equation as
\be
\rho = \frac{ \rho_{0}}{a^3}.
\ee

Variation of the relevant action with respect to $\phi$ yields the wave equation
\be
\ddot{\phi} + 3\frac{\dot{a}}{a}\dot{\phi} + V'(\phi) = {\omega}^2 \phi ,
\ee
~~~where a prime is a differentiation w.r.t. $\phi$. \\
 As the scalar field actually has two components, a third conservation equation is found as 
\be
\omega = \frac{A_{0}}{{\phi}^2 a^3}
\ee
$A_{0}$ being a constant.\\
  In the spintessence model described by Boyle et al \cite{boyle}, it has been considered that $\omega$ is a very slowly varying function of time which indeed can be considered as a constant
over the entire period of dust dominated era.
So for constant $\omega$, equation (7) yields

\be
{\phi}^2 a^3 = A ,
\ee
$A$ being a positive constant. \\

It deserves mention that only two equations amongst (5),
 (6) and (8) are independent as any one of them can be derived from the 
Einstein field equations with the help of the other two in view of the 
Bianchi identities. So, we have four unknowns, namely, $a$, $\rho$, $\phi$ and 
$V(\phi)$ and four equations, e.g, (3), (4) and two from (5)-(8). So this is 
a determined problem and an exact solution is on cards without any input. From equations (3), (4) and (5) one can write
\be
\frac{\ddot{a}}{a} - \frac{ \dot{a}^2}{a^2} = -\frac{\rho_{0}}{2 a^3} - \frac{1}{2} \dot{\phi}^2 -\frac{1}{2} {\omega}^2 {\phi}^2 
\ee
which takes the form
\be
a^3 \frac{dH}{dt} = -l -m H^2 ,
\ee
where $H = \frac{\dot{a}}{a}$, the Hubble parameter, \\
      $l = \frac{1}{2}(\rho_{0} + {\omega}^2 A)$ and
      $m = \frac{9 A}{8}$ are positive constants.

~ In deriving this, the equation (8) has been used to eliminate $\phi$ and 
$\dot{\phi}$ in terms of $a$ and $\dot{a}$. Now we make a transformation 
of time coordinate by the equation
\be
\frac{dx}{dt} = \frac{1}{a^3}.
\ee

\par As $\frac{dx}{dt}$ is positive definite ( $a$ is the scale factor and 
cannot take negative values ), we find that $x$ is a monotonically increasing 
function of $t$. So one can use $x$ as the new cosmic time without any loss 
of generality and deformation of the description of events. \\
In terms of $x$, equation (10) can be written as
\be
\frac{dH}{dx} = -l -m H^2 ,
\ee
which can be readily integrated to yield
\be
H = n ~ tan(\beta - mnx),
\ee
where $n^2 = \frac{l}{m}$, is a positive constant, and $\beta$ is a constant 
of integration.
\par Now we use this $H$ as a function of the new cosmic time variable $x$ and find out the behaviour of the deceleration parameter $q$, which is defined as
\be
q = -\frac{\dot{H}}{H^2} - 1 =  -\frac{H^\dagger}{a^3H^2} - 1
\ee 
where a dagger indicates differentiation w.r.t. $x$.\\
Equation (12) now yields
\be
q = -1 +\frac{1}{a^3} \left( \frac{l}{H^2} + m \right) ~.
\ee

\par Both $m$ and $l$ are positive constants. So $q$ has a `zero', when
\be
{a_{1}}^3 = \frac{l}{H_{1}^2} + m~.
\ee
The suffix 1 indicates the values of the quantities for $q = 0$. Furthermore, 
equation (15) can be differentiated to yield 
\be
{\frac{dq}{da}~\vline}_{~1} = -\frac{1}{a_{1}^4}\left
     [\frac{l}{H_{1}^2} + 3m\right]~,
\ee
at the point $q = 0$, $a = a_{1}$ and $H = H_{1}$. In deriving the 
equation (17), equations (10) and (16) have been used. The last equation 
clearly shows that $q$ is a decreasing function of $a$ atleast when $q = 0$. 
So $q$ definitely 
enters a negative value regime from a positive value at $a = a_{1}$ and 
$H = H_{1}$. 
\par For the sake of completeness, the solution for the scale factor $a$ 
can be found out by integrating equation (13) as 
\be
a =\left[{\frac{m}{ ln\vline\frac{1}
            {cos( \beta - mnx )}\vline^3}}\right]^{1/3}~.
\ee
\par So evidently, the spintessence model proposed by Boyle et al \cite{boyle} 
passes the `fitness test', the deceleration parameter enters into a negative 
value in a ``finite past". In view of the high degree of non linearity of 
Einstein equations, exact analytic solutions are indeed more dependable, and 
the present investigation provides that in support of a spintessence model. 
Also, the constants of the theory ( such as $\omega$ ) and the constants of 
integration ( such as $l$ and $m$ ) are still free parameters and hence 
provides the `comfort zone' for fitting into the observational results. 
Another feature of this study is that the results obtained are completely 
independent of the choice of potential $ V = V(\phi)$. This feature provides a 
bonus, as different potentials, used as the quintessence matter, are hardly 
well-motivated and do not have any proper physical background. Recently quite 
a few investigations show that a complex scalar field indeed serves 
the purpose of driving a late time acceleration \cite{je}. But most of these 
investigations either invoke the solution in some limit ( such as for a 
large $a$ ) or use some tuning of the form of the potential. The present 
investigation provides a better footing for them as it 
shows the transition of $q$ analytically. It also deserves mention that  Bento,
Bertolami and Sen \cite{anjan} showed the effectiveness of a Chaplygin gas 
as a quintessence matter. It is interesting to note that under some 
assumptions this kind of a fluid formally resembles a complex scalar field 
as discussed in this work. 

\vskip .2in

\noindent
{\bf Acknowledgement}

\vskip .1in
This work had been partially supported by DAE (India). The authors are grateful to Anjan Ananda Sen for pointing out the existence of the conservation equation (7).


\vskip .2in
\begin{thebibliography}{25}
\bibitem{perl}S. Perlmutter et al, Nature, {\bf391}, 51 (1998);\\
               S. Perlmutter et al, Astrophys. J., {\bf483}, 565 (1997);\\
               S. Perlmutter et al, ibid, {\bf517}, 565 (1999);\\
\bibitem{gar}  P. M. Garnavich et al, Astrophys. J., {\bf 509}, 74 (1998);\\
               A. G. Reiss et al, Astrophys. J., {\bf 116}, 1009 (1998).
\bibitem{varun} V. Sahni and A. Starobinsky, Int. J. Mod. Phys. D, {\bf 9}, 373 (2000);\\
            V. Sahni, Class. Quantum. Grav., {\bf 19}, 3435 (2002);\\
            T. Padmanabhan, Phys. Rep. (in press), hep-th/0212290.\\
            S. M. Carroll, {\it Carnegie Observatories Astrophysics series, Vol. 2, Measuring and Modelling the universe},~~editor W. L. Freedman ( Cambridge : Cambridge University Press, 2003 ).     
\bibitem{padma}T. Padmanabhan and T. Roy Chowdhury, Mon. Not. R. Astron. Soc., (in
press), \\~~~~~~~~~~~~~~~~~~~~~~~~~~~~astro-ph/0212573.\\
\bibitem{amen}L. Amendola, Mon. Not. R. Astron. Soc., {\bf 342}, 221 (2003).\\
\bibitem{riess}A. G. Reiss, astro-ph/0104455.\\
\bibitem{boyle}L. A. Boyle, R. R. Caldwell, M. Kamionkowski, Phys. Lett. B, {\bf 545}, 17 (2002).\\
\bibitem{vilen}A. Vilenkin, Phys. Rep. {\bf 121}, 263 (1985).\\
\bibitem{je}Je - An Gu and W - Y. P. Hwang, astro-ph/0105099;\\
            Y. H. Wei and Y. Z. Zhang, astro-ph/0402515;\\
            A. V. Yurov, hep-th/0208129. 
\bibitem{anjan}M. C. Bento, O. Bertolami, A. A. Sen, Phys. Rev. D, {\bf 66}, 043507 (2002).\\



\end{thebibliography}
\end{document}